\documentclass[twocolumn,showpacs,prb]{revtex4-1}%
\usepackage{amsfonts}
\usepackage{amsmath}
\usepackage{amssymb}
\usepackage{subfigure}
\usepackage{graphicx}
\usepackage{txfonts}%
\providecommand{\U}[1]{\protect\rule{.1in}{.1in}}

\begin{document}

\title{Oblique Klein tunneling in 8-\textit{Pmmn} borophene \textit{p-n} junctions}
\author{Shu-Hui Zhang$^{1}$}
\email{shuhuizhang@mail.buct.edu.cn}
\author{Wen Yang$^{2}$}
\email{wenyang@csrc.ac.cn}
\affiliation{$^{1}$College of Science, Beijing University of Chemical Technology, Beijing,
100029, China}
\affiliation{$^{2}$Beijing Computational Science Research Center, Beijing 100193, China}

\begin{abstract}
The 8-\textit{Pmmn} borophene is one kind of new elemental monolayer, which hosts
anisotropic and tilted massless Dirac fermions (MDF). The planar \textit{p-n}
junction (PNJ) structure as the basic component of various novel devices based
on the monolayer material has attracted increasing attention. Here, we
analytically study the transport properties of anisotropic and tilted MDF
across 8-\textit{Pmmn} borophene PNJ. Similar to the isotropic MDF across
graphene junctions, perfect transmission exists but its direction departures
the normal direction of borophene PNJ induced by the anisotropy and tilt,
i.e., oblique Klein tunneling. The oblique Klein tunneling does not depend on
the doping levels in \textit{N} and \textit{P} regions of PNJ as the normal Klein tunneling but
depends on the junction direction. Furthermore, we analytically derive the
special junction direction for the maximal difference between perfect
transmission direction and the normal direction of PNJ and clearly
distinguish the respective contribution of anisotropy and tilt underlying the
oblique Klein tunneling. In light of the rapid advances of experimental
technologies, we expect the oblique Klein tunneling to be observable in the near future.

\end{abstract}
\maketitle

\section{Introduction}

Graphene was the first atomically thin two-dimensional layer, and it hosts the
relativistic massless Dirac fermions (MDF) which possesses various unique
physics and possible applications \cite{CastroRMP2009}. Following the seminal
discovery of graphene, great efforts have been paid to search for new Dirac
materials which can host MDF \cite{Wehling2014,WangNSR2015}, especially in
monolayer structures. Boron is a fascinating element due to its chemical and
structural complexity, and boron-based nanomaterials of various dimensions have
attracted a lot of attention \cite{C7CS00261K,Kondo2017}, where the
two-dimensional phases of boron with space groups \textit{Pmmm} and
\textit{Pmmn}, hosting MDF, were also theoretically predicted
\cite{PhysRevLett.112.085502}. As one of the most stable predicted structures,
the two-dimensional phase of \textit{Pmmn} boron (named 8-\textit{Pmmn}
borophene) was studied in detail and its unprecedented electronic properties
were revealed by first-principles calculations \cite{PhysRevB.93.241405}. The tight-binding model of 8-\textit{Pmmn} borophene was developed
\cite{PhysRevB.94.165403,PhysRevB.97.125424} and an effective low-energy
Hamiltonian in the vicinity of Dirac points was proposed based on symmetry
consideration, and the pseudomagnetic field was also predicted similar to the
strained graphene \cite{PereiraPRL2009,Naumis2017}. In 8-\textit{Pmmn}
borophene, the effective low-energy Hamiltonian was used to study the plasmon
dispersion and screening properties by calculating the density-density
response function \cite{PhysRevB.96.035410}, the optical conductivity
\cite{PhysRevB.96.155418}, and Weiss oscillations \cite{PhysRevB.96.235405}.
The fast growing experimental confirmation of various borophene monolayers
\cite{MannixS1513,FengNC2016,PhysRevLett.118.096401} make 8-\textit{Pmmn}
borophene very promising.

The 8-\textit{Pmmn} borophene is one kind of elemental two-dimensional
material and hosts MDF \cite{PhysRevLett.112.085502}, whose high mobility
\cite{C7CP03736H} promises its future device applications in electronic and
electron optics. The planar \textit{p-n} junction (PNJ) structure is the
basic component of various novel devices, in which MDF exhibits a lot of
exotic properties with Klein tunneling as an example. Klein tunneling
\cite{KleinZP1929} is one phenomenon in quantum electrodynamics implying the
unimpeded penetration (i.e., perfect tunneling) of normally incident
relativistic particles regardless of the height and width of potential
barriers. Klein tunneling was firstly introduced into graphene
\cite{KatsnelsonNatPhys2006} and there are extensive theoretical
\cite{CheianovPRB2006,PereiraPRB2006,BaiPRB2007,BeenakkerPRB2008,SetareJPCM2010,RoslyakJPCM2010,ZebPRB2008,SoninPRB2009,SchelterPRB2010,YangPRB2011c,RozhkovPhysRep2011,LiuPRB2012b,GiavarasPRB2012,RodriguezJAP2012,PopoviciPRB2012,HeinischPRB2013,LogemannPRB2015,BaiYangChang2016,OhPRL2016,ErementchoukJPCM2016,DowningJPCM2017,PhysRevB.97.035420}
and experimental
\cite{HuardPRL2007,GorbachevNL2008,StanderPRL2009,YoungNatPhys2009,RossiPRB2010,SajjadPRB2012,SutarNL2012,RahmanAPL2015,GutierrezNatPhys2016,ChenScience2016,LaitinenPRB2016,BaiPRB2017}
 and application studies
\cite{SajjadAPL2011,JangPNAS2013,Wilmart2DM2014,ChenPRB2015,SousaJAP2017}. In
contrast to the isotropic MDF in graphene, the MDF in 8-\textit{Pmmn}
borophene is anisotropic and tilted, so the new feature for Klein tunneling is
expected. In fact, two recent works have reported the oblique Klein tunneling
(i.e., the perfect transmission direction does not overlap with the normal
direction of PNJ) induced by the anisotropy of two-dimensional MDF
\cite{li2017} and the tilt of three-dimensional MDF \cite{PhysRevB.97.235113},
respectively. Thus, 8-\textit{Pmmn} borophene provides an ideal platform to
study Klein tunneling in the presence of interplay between anisotropy and tilt
of two-dimensional MDF.

In the present paper, we study analytically the transmission properties of
anisotropic and tilted MDF across 8-\textit{Pmmn} borophene PNJ. The
anisotropy and tilt together lead to the oblique Klein tunneling, which does
not depend on the doping levels in \textit{N} and \textit{P} regions of PNJ as the normal Klein
tunneling but depends on the junction direction. There is a special junction
direction for the maximal difference between the perfect transmission
direction and the normal direction of PNJ, which is obtained analytically. The
respective contribution of anisotropy and tilt to the oblique Klein tunneling
is also distinguished, which is useful to identify the nature of energy
dispersion. The rest of this paper is organized as follows. In Sec. II, we
introduce two coordinate systems for the 8-\textit{Pmmn} borophene PNJ, present the
intrinsic electronic properties of 8-\textit{Pmmn} borophene, and the detailed
derivation of transmission of anisotropic tilted MDF across PNJ. In Sec. III,
we demonstrate analytically the existence of perfect transmission and show
the noncollinear nature of group velocities and momenta of incident states
induced by the anisotropy and/or tilt leading to the oblique Klein tunneling.
Finally, we give a brief summary in Sec. IV.

\section{Theoretical formalism}

\begin{figure}[ptbh]
\includegraphics[width=\columnwidth,clip]{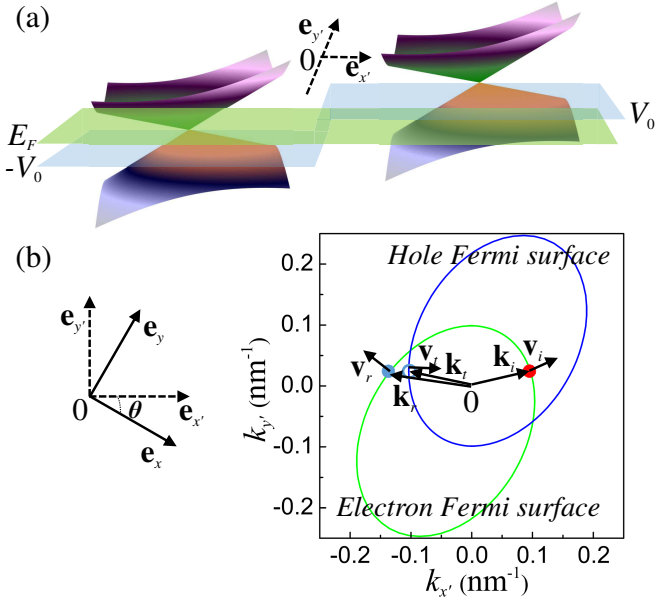}\caption{(a) 8-Pmmn
borophene \textit{p-n} junction. The normal (tangential) direction of the junction
interface defines the $x^{\prime}$ ($y^{\prime}$) axis of the Cartesian coordinate
system $x^{\prime}-y^{\prime}$. Along the energy axis, the Dirac point of \textit{N}
(\textit{P}) region is at $-V_{0}$ ($V_{0}$) and the aligned Fermi level is at $E_{F}$.
(b) Two Cartesian coordinate systems $x-y$ and $x^{\prime}-y^{\prime}$ are
introduced and they are rotated to each other with $\theta$, so $\theta$ can
be regarded as the junction direction. Using the coordinate systems, the
electron (hole) Fermi surface in \textit{N} (\textit{P}) region is plotted. On the Fermi
surfaces, we also plot the noncollinear momenta and group velocities of the
incident, reflection, and transmission states. Here, we use $\theta=\pi/6$ for
the junction direction, and $\varepsilon_{N}=\varepsilon_{P}=0.04$ eV for the
doping levels in N and P regions.}%
\label{structure}%
\end{figure}

The 8-\textit{Pmmn} borophene PNJ structure is shown schematically by Fig.
\ref{structure}(a) and it has the left \textit{N} region and right \textit{P} region. For the
PNJ, the Cartesian coordinate system $x^{\prime}-y^{\prime}$ is introduced, in
which $x^{\prime}$ ($y^{\prime}$) axis is along the normal
(tangential)\ direction of junction interface. The Hamiltonian of PNJ in Fig.
\ref{structure}(a) has the form:%

\begin{equation}
\hat{H}=(\hat{H}_{0}-V_{0})\Theta(-x^{\prime})+(\hat{H}_{0}+V_{0}%
)\Theta(x^{\prime}),
\end{equation}
where $-V_{0}$ ($V_{0}$) is the gate-induced scalar potential in the \textit{N} (\textit{P}) region
by assuming $V_{0}>0$ without loss of generality, and $\Theta(x^{\prime})$ is
the step function: $\Theta(x^{\prime})=1$ for $x^{\prime}>0$ and
$\Theta(x^{\prime})=0$ for $x^{\prime}<0$. The Fermi level $E_{F}$ determines
the doping level in the \textit{N} (\textit{P}) region as $\varepsilon_{N}\equiv E_{F}+V_{0}$
($\varepsilon_{P}\equiv E_{F}-V_{0}$), where a positive (negative) doping
level means electron or \textit{N} (hole or \textit{P}) doping, so $\varepsilon_{N}>0$ and
$\varepsilon_{P}<0$ in our case. For the intrinsic Hamiltonian $\hat{H}_{0}$
of 8-Pmmm borophene, we introduce the Cartesian coordinate system $x-y$ which
is rotated in terms of $\theta$ relative to the coordinate system $x^{\prime
}-y^{\prime}$ as shown in Fig. \ref{structure}(b), so $\theta$ can be used to
indicate the junction direction. The transformation relation between the basis
vectors $(\mathbf{e}_{x},\mathbf{e}_{y})$\ and $(\mathbf{e}_{x^{\prime}%
},\mathbf{e}_{y^{\prime}})$ of two coordinate systems is%

\begin{equation}%
\begin{bmatrix}
\mathbf{e}_{x}\\
\mathbf{e}_{y}%
\end{bmatrix}
=\left[
\begin{array}
[c]{cc}%
\cos\theta & -\sin\theta\\
\sin\theta & \cos\theta
\end{array}
\right]
\begin{bmatrix}
\mathbf{e}_{x^{\prime}}\\
\mathbf{e}_{y^{\prime}}%
\end{bmatrix}
. \label{CSR}%
\end{equation}
As a result, an arbitrary vector can be denoted by $\mathbf{O=(}O_{x}%
,O_{y}\mathbf{)}$ in the coordinate system $x-y$ and by $\mathbf{O}^{\prime
}\mathbf{=(}O_{x^{\prime}},O_{y^{\prime}}\mathbf{)}$ in the coordinate system
$x^{\prime}-y^{\prime}$, and the vector's components in two coordinate systems
are related to each other:%

\begin{subequations}
\label{VE}%
\begin{align}
O_{x}  &  =O_{x^{\prime}}\cos\theta-O_{y^{\prime}}\sin\theta,\\
O_{y}  &  =O_{x^{\prime}}\sin\theta+O_{y^{\prime}}\cos\theta.
\end{align}
Obviously, $O=O^{\prime}$ with $O=\left\vert \mathbf{O}\right\vert $ and
$O^{\prime}=\left\vert \mathbf{O}^{\prime}\right\vert $.

\subsection{The intrinsic electronic properties of 8-\textit{Pmmn} borophene}

The Hamiltonian of anisotropic tilted MDF around one Dirac point of 8-Pmmn
borophene is given
by\cite{PhysRevB.94.165403,PhysRevB.96.035410,PhysRevB.96.235405}%

\end{subequations}
\begin{equation}
\hat{H}_{0}=v_{x}\sigma_{x}\hat{p}_{x}+v_{y}\sigma_{x}\hat{p}_{y}%
+v_{t}\mathbf{I}_{2\times2}\hat{p}_{y},
\end{equation}
where $\hat{p}_{x,y}$ are the momentum operators, $\sigma_{x,y}$ are
$2\times2$ Pauli matrices, and $\mathbf{I}_{2\times2}$ is the $2\times2$ identity
matrix. Throughout this paper, we assume $\hbar\equiv1$. The anisotropic
velocities are $v_{x}=0.86v_{F}$, $v_{y}=0.69v_{F}$, and $v_{t}=0.32v_{F}$
with $v_{F}=10^{6}$ m/s. The energy dispersion and the corresponding wave
functions of $\hat{H}_{0}$\ are, respectively,%

\begin{equation}
E_{\lambda,\mathbf{k}}=v_{t}k_{y}+\lambda v_{x}\sqrt{k_{x}^{2}+\gamma_{1}%
^{2}k_{y}^{2}}, \label{EV}%
\end{equation}
and%
\begin{equation}
\psi_{\lambda,\mathbf{k}}(\mathbf{r})=\frac{1}{\sqrt{2}}%
\begin{bmatrix}
1\\
\lambda\frac{\left(  k_{x}+i\gamma_{1}k_{y}\right)  }{\sqrt{k_{x}^{2}%
+\gamma_{1}^{2}k_{y}^{2}}}%
\end{bmatrix}
e^{i\mathbf{k}\cdot\mathbf{r}}. \label{WF}%
\end{equation}
Here, $\gamma_{1}=v_{y}/v_{x}$, $\lambda=+$ ($\lambda=-$)\ denotes the
conduction (valence) band, $\mathbf{k}=(k_{x},k_{y})$ is the momentum, and
$\mathbf{r}=(x,y)$ is the position vector. The azimuthal angle of $\mathbf{k}$
relative to the $x$ axis is $\phi_{\mathbf{k}}$ which leads to tan$\phi
_{\mathbf{k}}=k_{y}/k_{x}$, so the energy dispersion of Eq. (\ref{EV}) becomes%

\begin{equation}
E_{\lambda,\mathbf{k}}=v_{t}k\sin\phi_{\mathbf{k}}+\lambda v_{x}k\sqrt
{\cos^{2}\phi_{\mathbf{k}}+\gamma_{1}^{2}\sin^{2}\phi_{\mathbf{k}}}.
\label{EVP}%
\end{equation}
To determine the shape of Fermi surface for the fixing energy $E_{\lambda
,\mathbf{k}}$, we can change Eq. (\ref{EV}) into
\begin{equation}
\frac{k_{x}^{2}}{k_{\lambda,a}^{2}}+\frac{(k_{y}+k_{\lambda,s})^{2}%
}{k_{\lambda,b}^{2}}=1, \label{ED1}%
\end{equation}
where
\begin{align*}
k_{\lambda,s}  &  =\frac{E_{\lambda,\mathbf{k}}v_{t}}{v_{y}^{2}-v_{t}^{2}},\\
k_{\lambda,a}^{2}  &  =\frac{E_{\lambda,\mathbf{k}}^{2}v_{y}^{2}}{v_{x}%
^{2}(v_{y}^{2}-v_{t}^{2})},\\
k_{\lambda,b}^{2}  &  =\frac{E_{\lambda,\mathbf{k}}^{2}v_{y}^{2}}{(v_{y}%
^{2}-v_{t}^{2})^{2}}.
\end{align*}
Clearly, Eq. (\ref{ED1}) is one equation of a shifted ellipse originated from
the anisotropic tilted electronic properties. On one hand, the ratio of the
semimajor and semiminor axes of the ellipse are $k_{\lambda,b}/k_{\lambda
,a}=v_{x}/\sqrt{(v_{y}^{2}-v_{t}^{2})}\approx\sqrt{2}$ which does not depend
on the energy dispersion $E_{\lambda,\mathbf{k}}$ (with the band index
$\lambda$)\ and\ is a constant determined only by the anisotropic velocities.
On the other hand, the ellipse is shifted along the $y$ axis such that its
center lies at $(0,k_{\lambda,s})$. Because $k_{\lambda,s}\propto
E_{\lambda,\mathbf{k}}$ and $E_{+,\mathbf{k}}\ $and $E_{-,\mathbf{k}}$ have
the opposite sign, the electron and hole Fermi surfaces are shifted oppositely
along the $y$ axis. Combing these two features, the electron Fermi surface in the
\textit{N} region and the hole Fermi surface in the \textit{P} region are shown in Fig. \ref{structure}(b).

For convenience, we can define the pseudospin vector $\mathbf{s}=(s_{x}%
,s_{y})$ for each state and%

\begin{subequations}
\label{PSC}%
\begin{align}
s_{x}  &  \equiv\langle\psi_{\lambda,\mathbf{k}}|\sigma_{x}|\psi
_{\lambda,\mathbf{k}}\rangle=\frac{\lambda k_{x}}{\sqrt{k_{x}^{2}+\gamma
_{1}^{2}k_{y}^{2}}},\\
s_{y}  &  \equiv\langle\psi_{\lambda,\mathbf{k}}|\sigma_{y}|\psi
_{\lambda,\mathbf{k}}\rangle=\frac{\lambda\gamma_{1}k_{y}}{\sqrt{k_{x}%
^{2}+\gamma_{1}^{2}k_{y}^{2}}}.
\end{align}
So the azimuthal angle $\phi_{\mathbf{s}}$ of $\mathbf{s}$ satisfies
\end{subequations}
\begin{equation}
\tan\phi_{\mathbf{s}}=s_{y}/s_{x}=\gamma_{1}\tan\phi_{\mathbf{k}}. \label{PSA}%
\end{equation}
Then, Eq. \ref{WF}(b) becomes%

\begin{equation}
\psi_{\lambda,\mathbf{k}}(\mathbf{r})=\frac{1}{\sqrt{2}}%
\begin{bmatrix}
1\\
e^{i\phi_{\mathbf{s}}}%
\end{bmatrix}
e^{i\mathbf{k}\cdot\mathbf{r}}.
\end{equation}

\subsection{Transmission of anisotropic tilted MDF across PNJ}

Firstly, because the translation invariance symmetry along the junction
interface of PNJ requires the conservation of the tangential momentum
$k_{y^{\prime}}$, it is convenient to derive the transmission probability of
anisotropic tilted MDF across PNJ in the coordinate system $x^{\prime
}-y^{\prime}$. Without loss of generality to consider the electron state
incident from the left \textit{N} region of PNJ, there are incident and reflection
states in \textit{N} region and transmission state in \textit{P} region. In the following, we
use
\begin{equation}
\psi^{\alpha}(\mathbf{r}^{\prime})=\frac{1}{\sqrt{2}}%
\begin{bmatrix}
1\\
e^{i\phi_{\mathbf{s}}^{\alpha}}%
\end{bmatrix}
e^{i\mathbf{k}_{\alpha}^{\prime}\cdot\mathbf{r}^{\prime}}%
\end{equation}
to denote the incident state $(\alpha=i)$, reflection state $(\alpha=r)$, and
transmission state $(\alpha=t)$ in the mixed coordinate systems. Here,
corresponding to the $\alpha$ state, $\phi_{\mathbf{s}}^{\alpha}$ is the azimuthal
angle of pseudospin vector in the coordinate system $(x,y)$, and
$\mathbf{k}_{\alpha}^{\prime}=(k_{\alpha,x}^{\prime},k_{y}^{\prime})$ is the
momentum in the coordinate system $x^{\prime}-y^{\prime}$. Note that we use
$k_{y}^{\prime}$ instead of $k_{\alpha,y}^{\prime}$ because $k_{y}^{\prime}$
is a conserved quantity. To solve the transmission problem, we obtain the
matching equation for three states across the PNJ at $y^{\prime}=0$:%

\begin{equation}
\psi^{i}(\mathbf{r}^{\prime})+r\psi^{r}(\mathbf{r}^{\prime})=t\psi
^{t}(\mathbf{r}^{\prime}).
\end{equation}
As a result, the reflection coefficient $r$\ and transmission coefficient $t$ are:%

\begin{subequations}
\label{SE}%
\begin{align}
r  &  =-\frac{e^{i\phi_{\mathbf{s}}^{i}}-e^{i\phi_{\mathbf{s}}^{t}}}%
{e^{i\phi_{\mathbf{s}}^{r}}-e^{i\phi_{\mathbf{s}}^{t}}},\\
t  &  =1-r.
\end{align}
The transmission probability of anisotropic tilted MDF across the PNJ is
\end{subequations}
\begin{equation}
T=1-|r|^{2}, \label{TP}%
\end{equation}
whose calculation requires the value of $\mathbf{k}_{\alpha}=(k_{\alpha
,x},k_{\alpha y})$ due to the Eqs. (\ref{PSC}) and (\ref{PSA}) for pseudospin
vector in the coordinate system $(x,y)$. Note that $k_{\alpha y}$ is not conserved.

Secondly, we show how to obtain $\mathbf{k}_{r}$ and $\mathbf{k}_{t}$ for the
calculation of $T$ by taking full advantage of the conserved $k_{y^{\prime}}$.
The incident state has the momentum $\mathbf{k}_{i}=(k_{i,x},k_{i,y}%
)=(k_{i}\cos\phi_{\mathbf{k}}^{i},k_{i}\sin\phi_{\mathbf{k}}^{i})$, which is
one given quantity, and Eq. (\ref{EVP}) implies
\begin{equation}
k_{i}=\frac{\varepsilon_{N}}{v_{t}\sin\phi_{\mathbf{k}}^{i}+v_{x}\sqrt
{\cos^{2}\phi_{\mathbf{k}}^{i}+\gamma_{1}^{2}\sin^{2}\phi_{\mathbf{k}}^{i}}}.
\end{equation}
This makes $\mathbf{k}_{i}^{\prime}=(k_{i,x^{\prime}},k_{y^{\prime}}%
)=(k_{i}\cos\phi_{\mathbf{k}^{\prime}}^{i},k_{i}\sin\phi_{\mathbf{k}^{\prime}%
}^{i})$ with $\phi_{\mathbf{k}^{\prime}}^{i}=\phi_{\mathbf{k}}^{i}-\theta$ by
considering Eq. (\ref{CSR}) for transformation relation between two coordinate
systems. Since $k_{y^{\prime}}=k_{i}\sin\phi_{\mathbf{k}^{\prime}}^{i}$ is
obtained and conserved, in the following, we need to derive $k_{r,x^{\prime}}$
and $k_{t,x^{\prime}}$. By using Eq. (\ref{VE}), we obtain%

\begin{subequations}
\label{CT}%
\begin{align}
k_{\alpha,x}  &  =k_{\alpha,x^{\prime}}\cos\theta-k_{y^{\prime}}\sin\theta,\\
k_{\alpha,y}  &  =k_{\alpha,x^{\prime}}\sin\theta+k_{y^{\prime}}\cos\theta.
\end{align}
Because $\mathbf{k}_{\alpha}$ satisfies Eq. (\ref{EV}) for energy dispersion, we obtain%

\end{subequations}
\begin{equation}
Ak_{\beta,x^{\prime}}^{2}+B_{\beta}k_{\beta,x^{\prime}}+C_{\beta}=0,
\label{QEOU}%
\end{equation}
where
\begin{align*}
A  &  =\cos^{2}\theta+(\gamma_{1}^{2}-\gamma_{2}^{2})\sin^{2}\theta,\\
B_{\beta}  &  =2\epsilon_{\beta}\gamma_{2}\sin\theta-2(1-\gamma_{1}^{2}%
+\gamma_{2}^{2})k_{y^{\prime}}\sin\theta\cos\theta,\\
C_{\beta}  &  =k_{y^{\prime}}^{2}(\sin^{2}\theta+\gamma_{1}^{2}\cos^{2}%
\theta-\gamma_{2}^{2}\cos^{2}\theta)+2\epsilon_{\beta}\gamma_{2}k_{y^{\prime}%
}\cos\theta-\epsilon_{\beta}^{2}.
\end{align*}
Here, $\gamma_{2}=v_{t}/v_{x}$, $\epsilon_{\beta}=\varepsilon_{\beta}/v_{x}$
with $\beta=N,P$. Equation (\ref{QEOU}) is a quadratic equation with one unknown and
has two formal roots:
\[
k_{\beta,x^{\prime}}^{\pm}=\frac{-B_{\beta}\pm\sqrt{B_{\beta}^{2}-4AC_{\beta}%
}}{2A}.
\]
The right-going and left-going eigenstates should be finite when $x^{\prime
}\rightarrow+\infty$ and $x^{\prime}\rightarrow-\infty$; this property can be
used to distinguish them. In order to distinguish the left-going and
right-going eigenstates, we make the replacement $E_{F}\rightarrow
E_{F}+i0^{+}$\ for the Fermi level in which $0^{+}$ is one positive infinitely
small quantity. As a result, the right-going $k_{i,x^{\prime}}=k_{N,x^{\prime
}}^{+}(=k_{i}\cos\phi_{\mathbf{k}^{\prime}}^{i})$ and $k_{t,x^{\prime}%
}=k_{P,x^{\prime}}^{+}$\ both have the positive infinitely small imaginary
part while the left-going $k_{r,x^{\prime}}=k_{N,x^{\prime}}^{-}$ has the
negative infinitely small imaginary part. To substitute $k_{y^{\prime}}$ and
derived $k_{r,x^{\prime}}$, $k_{t,x^{\prime}}$ into Eq. (\ref{CT}), one
obtains $\mathbf{k}_{r,t}$, then can calculate the transmission probability
$T$. For one given $\mathbf{k}_{i}$, we plot schematically $\mathbf{k}%
_{\alpha}$ on the electron and hole Fermi surfaces in Fig. \ref{structure}(b).

\section{Results and discussions}

In this section, we present the numerical results for the transmission
probability of anisotropic tilted MDF across the borophene PNJ and discuss
the underlying physics.

\subsection{Existence of perfect transmission}

\begin{figure}[ptbh]
\includegraphics[width=\columnwidth,clip]{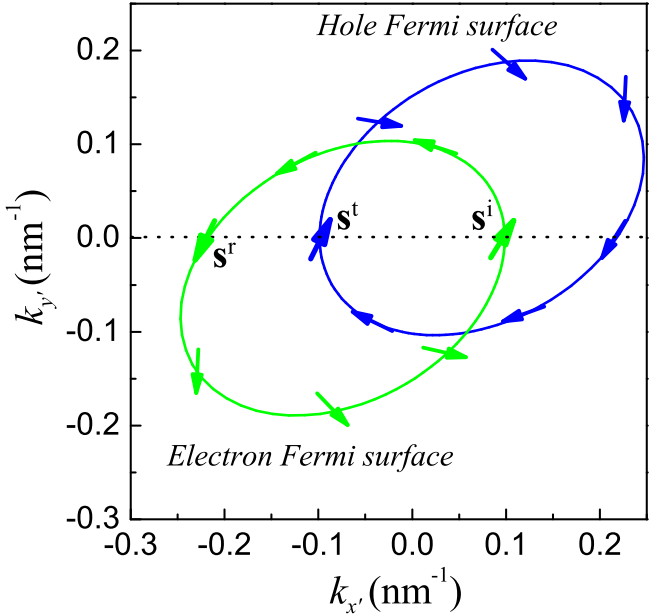}\caption{Pseudospin texture
on the electron (hole) Fermi surface denoted by the green (blue) circle in the \textit{N}
(\textit{P}) region. Perfect transmission occurs when $k_{y^{\prime}}=0$ (i.e., the
black dotted line), meanwhile, the pseudospin orientation of the incident state is
parallel to that of the transmission state and is antiparallel to that of the
reflection state. Here, we use $\theta=\pi/3$ for the junction direction and
$\varepsilon_{N}=\varepsilon_{P}=0.04$ eV for the doping levels in \textit{N} and \textit{P}
regions.}%
\label{spin}%
\end{figure}

Klein tunneling is one of the most exotic consequences of quantum
electrodynamics, which was firstly discussed in condensed matter physics
through investigating the tunneling properties of low-energy quasiparticles in
graphene junctions \cite{KatsnelsonNatPhys2006}. In graphene, the low-energy
quasiparticles are isotropic MDF, which implies perfect transmission of MDF
across PNJ at normal incidence. Klein tunneling is expected to occur in the
borophene PNJ though its MDF are anisotropic and tilted. Previous to the
numerical results, we analytically demonstrate the existence of perfect
transmission in the borophene PNJ. From (Eq. \ref{TP}), perfect transmission
(i.e., $T=1$) occurs when $r=0$. $r$ is given by Eq. \ref{SE}(a), $r=0$ leads
to $e^{i\phi_{\mathbf{s}}^{i}}=e^{i\phi_{\mathbf{s}}^{t}}$ and $e^{i\phi_{\mathbf{s}}^{r}}\neq
e^{i\phi_{\mathbf{s}}^{t}}$ which is just the well-known conservation of pseudospin
$\mathbf{s}^{\alpha}$ \cite{AllainEPJB2011}, i.e., $\mathbf{s}^{i}%
\shortparallel\mathbf{s}^{t}$. Recalling that Klein tunneling occurs for the
incident electron state with zero tangential momentum along the junction
interface of graphene junctions \cite{AllainEPJB2011}, in the borophene PNJ,
$k_{y^{\prime}}=0$ should be the necessary condition for the Klein tunneling.
As shown in Fig. \ref{structure}(b), the conserved $k_{y^{\prime}}$ determines
the momentum positions of $\mathbf{k}_{t}$ ($\mathbf{k}_{i}$, $\mathbf{k}_{r}%
$)\ on the hole (electron) Fermi surface. When $k_{y^{\prime}}=0$,
$\mathbf{k}_{i}$, $\mathbf{k}_{r}$, and $\mathbf{k}_{t}$ are collinear.
Referring to Eqs. (\ref{PSC}) and (\ref{PSA}) for the pseudospin vector, we obtain
$\phi_{\mathbf{s}}^{t}=\phi_{\mathbf{s}}^{i}$ and $\phi_{\mathbf{s}}^{r}%
=\phi_{\mathbf{s}}^{i}+\pi$, i.e., the pseudospin orientation of the incident
state is parallel to that of the transmission state and is antiparallel to that of the
reflection state as shown by Fig. \ref{spin}, so perfect transmission exists
for the incident anisotropic tilted MDF with $k_{y^{\prime}}=0$ or
$\phi_{\mathbf{k}^{\prime}}^{i}=\theta$\ in the borophene PNJ.

\subsection{Noncollinear features of $\mathbf{v}_{\alpha}$ and $\mathbf{k}%
_{\alpha}$ induced by anisotropy and tilt}

However, for a wave packet, the direction of center-of-mass motion and energy
flow are described by the group velocity instead of the momentum
\cite{li2017}. For isotropic MDF in graphene junctions, the group velocity and
momentum of each state are collinear, so normal Klein tunneling is obtained.
For the incident electronic state from the left \textit{N} region of borophene PNJ, the
group velocity $\mathbf{v}_{i}=(v_{i,x},v_{i,y})$ is determined by the energy
dispersion of the conduction band (i.e., $\lambda=1)$ and%

\begin{subequations}
\label{GV}%
\begin{align}
v_{i,x}  &  \equiv\frac{\partial E_{+,\mathbf{k}_{i}}}{\partial k_{i,x}}%
=\frac{v_{x}k_{i,x}}{\sqrt{k_{i,x}^{2}+\gamma_{1}^{2}k_{i,y}^{2}}},\\
v_{i,y}  &  \equiv\frac{\partial E_{+,\mathbf{k}_{i}}}{\partial k_{i,y}}%
=v_{t}+\frac{v_{x}\gamma_{1}^{2}k_{i,y}}{\sqrt{k_{i,x}^{2}+\gamma_{1}%
^{2}k_{i,y}^{2}}},
\end{align}
which leads to%
\end{subequations}
\begin{equation}
\tan\phi_{_{\mathbf{V}}}^{i}=\gamma_{1}^{2}\tan\phi_{\mathbf{k}}^{i}%
+\gamma_{2}\sqrt{1+\gamma_{1}^{2}\tan^{2}\phi_{\mathbf{k}}^{i}}. \label{GVA}%
\end{equation}
Here, $\phi_{_{\mathbf{V}}}^{i}$ is the azimuthal angle of $\mathbf{v}_{i}$.
From Eq. (\ref{GV}), the anisotropy and tilt both lead to the noncollinear
feature of the group velocity and the momentum. Similarly, one can define
$\mathbf{v}_{r,t}$. In Fig. \ref{structure}(b), we also plot $\mathbf{v}%
_{\alpha}$ to clearly show the noncollinear features of $\mathbf{v}_{\alpha}$
and $\mathbf{k}_{\alpha}$ which will bring about unique transmission
properties in the borophene PNJ.

\subsection{Oblique Klein tunneling}

\begin{figure*}[ptbh]
\includegraphics[width=2.0\columnwidth,clip]{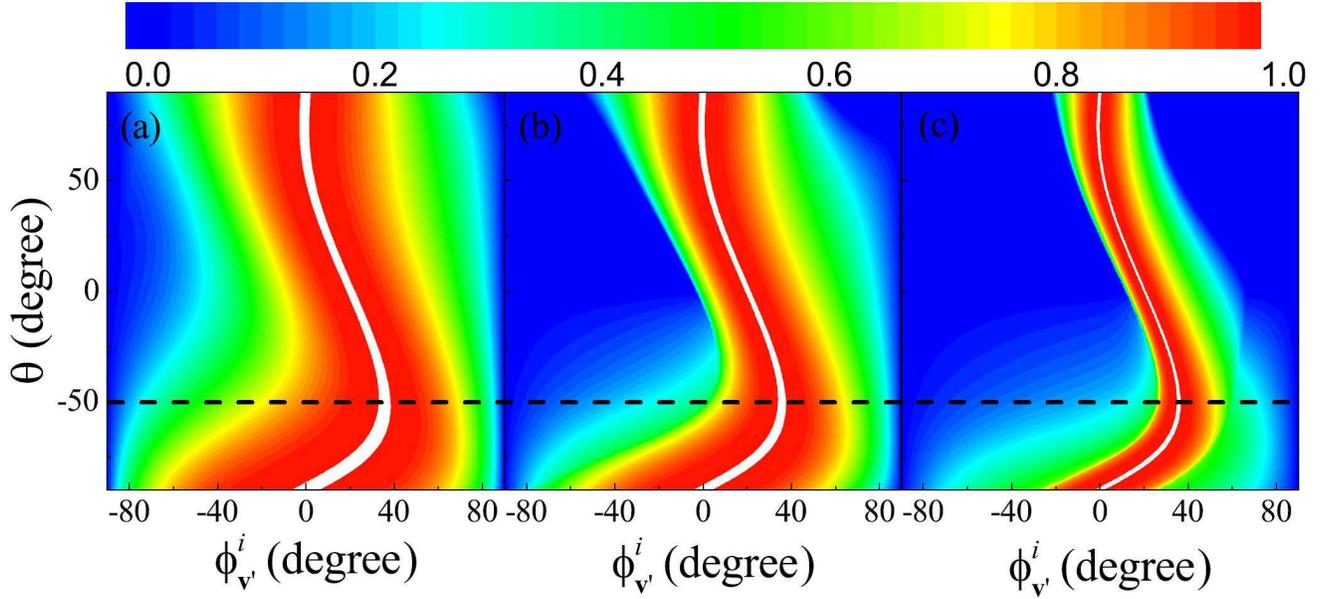}\caption{Contour plot
for transmission probability as the function of the junction direction
$\theta$ and the incident angle $\phi_{\mathbf{v}^{\prime}}^{i}=\phi
_{\mathbf{v}}^{i}-\theta$ relative to the normal direction of borophene
junction. (a) $E_{F}=-0.02$ eV, (b) $E_{F}=0$ eV, and (c) $E_{F}=0.02$ eV. In
each panel, $V_{0}=0.04$ eV and the white color region with the transmission
probability $T\geq0.999$ is used to denote the parameter points for perfect
transmission. The dashed black line denotes the special junction direction for
the oblique Klein tunneling with the maximal difference between the perfect
transmission direction and the normal direction of junction.}%
\label{tran}%
\end{figure*}

In the Cartesian coordinate system $x-y$, the electron and hole Fermi surfaces
have the mirror symmetry about the $y$ axis as shown in Fig. \ref{structure}%
(b), so the junction direction $\theta$ (i.e., the rotation angle between two
coordinate systems) can be limited into the angle range $[-90^{\circ
},90^{\circ}]$. Note that two cases for $\theta=90^{\circ
}$ and $\theta=-90^{\circ}$ are not equivalent because of the different matching conditions for the incident, reflection, and transmission states. The group velocity of the incident state relative to the normal
direction (i.e., $x^{\prime}$ axis)\ of borophene PNJ has the azimuthal angle
$\phi_{\mathbf{v}^{\prime}}^{i}=\phi_{\mathbf{v}}^{i}-\theta$. Due to the
isotropic nature of MDF in graphene junctions, perfect transmission occurs
when $\phi_{\mathbf{v}^{\prime}}^{i}=0$. The case is very different in the
borophene PNJ. As shown in Fig. \ref{tran}, we present the contour plot for
transmission probability of anisotropic tilted MDF as the function $\theta$
and $\phi_{\mathbf{v}^{\prime}}^{i}$ by considering different doping levels,
i.e., (a) $E_{F}=-0.02$ eV, (b) $E_{F}=0$, and (c) $E_{F}=0.02$ eV when $V_{0}=0.04$ eV.
Figure \ref{tran} shows several interesting properties. (1) As changing junction
direction of borophene PNJ by tuning $\theta$, perfect transmission must occur
consistent with the analytical demonstration in Sec. IIIA. (2) The direction
for perfect transmission deviates the normal direction of borophene PNJ, i.e.,
$\phi_{\mathbf{v}^{\prime}}^{i}\neq0$. For convenience, this phenomenon is
named as oblique Klein tunneling, which is in sharp contrast to the normal
Klein tunneling in the graphene PNJ. (3) By comparing the perfect
transmission in three subfigures,\ we find that the departure of oblique Klein
tunneling from the normal direction of borophene PNJ does not depend on the
doping level. The behavior can be understood in terms of Eq. (\ref{GVA}) in
which$\ \phi_{\mathbf{k}}^{i}=\theta$ for perfect transmission as discussed in
Sec. IIIA. As a result, the difference between the perfect transmission
direction and the normal direction only depends on $\theta$. Furthermore, the
special $\theta_{m}$ gives the maximal difference between the two directions. To
obtain the $\theta_{m}$ for oblique Klein tunneling, we need to maximize
$(\theta-\phi_{\mathbf{v}})$ or equivalently $\tan(\theta-\phi_{\mathbf{v}})$.
Using Eq. (\ref{GVA}), we obtain%

\begin{equation}
\tan(\theta-\phi_{\mathbf{v}})=\frac{1-\gamma_{1}^{2}}{\gamma_{2}}\cos
\tilde{\theta}-\frac{1+\gamma_{2}^{2}-\gamma_{1}^{2}}{\gamma_{2}}\frac
{\cos\tilde{\theta}}{1+\gamma\sin\tilde{\theta}},
\end{equation}
where $\gamma=\gamma_{2}/\gamma_{1}$ and
\begin{equation}
\tan\tilde{\theta}=\gamma_{1}\tan\theta. \label{ANGR}%
\end{equation}
The extrema of $\tan(\theta-\phi_{\mathbf{v}})$ is determined by
\[
\frac{\partial\tan(\theta-\phi_{\mathbf{v}})}{\partial\tilde{\theta}}=0,
\]
which gives
\begin{equation}
az^{3}+bz^{2}+cz+d=0. \label{CE}%
\end{equation}
Here, $z=\sin\tilde{\theta}$, $a=\gamma^{2}$, $b=2\gamma$, $c=1-c_{1}/c_{2}$ ,
and $d=-\gamma c_{1}/c_{2}$ with $c_{1}=(1+\gamma_{2}^{2}-\gamma_{1}%
^{2})/\gamma_{2}$ and $c_{2}=(1-\gamma_{1}^{2})/\gamma_{2}$. Equation (\ref{CE}) is
one cubic equation, and its general solutions are%

\[
z_{n}=-\frac{1}{3a}(b+\xi^{n}C+\frac{\Delta_{0}}{\xi^{n}C}),n\in\{0,1,2\},
\]
with%

\[
\xi=-\frac{1}{2}+\frac{\sqrt{3}}{2}i
\]
and%

\begin{align*}
\Delta &  =18abcd-4b^{3}d+b^{2}c^{2}-4ac^{3}-27a^{2}d^{2},\\
\Delta_{0}  &  =b^{2}-3ac,\\
\Delta_{1}  &  =2b^{3}-9abc+27a^{2}d,\\
C  &  =\sqrt[3]{\frac{\Delta_{1}\pm\sqrt{-27a^{2}\Delta}}{2}}.
\end{align*}
Because $\Delta\simeq2.617>0$, the cubic equation has three real roots and
they are%

\begin{align*}
z_{0}  &  \approx-4.565,\\
z_{1}  &  \approx-0.694,\\
z_{2}  &  \approx0.946.
\end{align*}
Recalling $\left\vert z\right\vert \leq1$ due to $z=\sin\tilde{\theta}$, so
$z_{1,2}$ are the two proper roots. Equation (\ref{ANGR}) gives
\begin{equation}
\theta=\arctan(\frac{\gamma_{1}z}{\sqrt{1-z^{2}}}). \label{ANGL}%
\end{equation}
Substituting $z_{1,2}$ into the above Eq. (\ref{ANGL}), we obtain
\begin{align*}
\theta_{1}  &  =-0.877\text{ rad}=-50.228^{\circ},\\
\theta_{2}  &  =1.302\text{ rad}=74.627^{\circ}.
\end{align*}
For the difference between the perfect transmission direction and the normal
direction, $\theta_{1}$ ($\theta_{2}$)\ gives the global maximum (local
minimum) as shown by Fig. \ref{tran}, so
\begin{equation}
\theta_{m}=\theta_{1}=-50.228^{\circ}%
\end{equation}
which is denoted\ by the black dashed line and is independent on doping level.
At $\theta_{m}$, it is $35.\,\allowbreak838^{\circ}$ for the maximal
difference between the perfect transmission direction and the normal
direction. The anisotropy and tilt of MDF in borophene PNJ both contribute to
$\theta_{m}$. If no tilt ($v_{t}=0$), $\theta_{m}$ contributed by the
anisotropy is $\theta_{m}^{a}=\arctan(1/\gamma_{1})=51.\,\allowbreak
259^{\circ}$. The drastic difference between $\theta_{m}^{a}$ and $\theta
_{m}$ implies that the tilt plays an important role leading to the oblique
Klein tunneling. Therefore, the unique features of oblique Klein tunneling are
intimately related to the nature of MDF, so they can be used to identify the
anisotropy and/or tilt of energy dispersion.

\section{Conclusions and outlook}

In this study, we investigate analytically the transport properties of
anisotropic and tilted MDF in the 8-\textit{Pmmn} borophene PNJ. The unique
oblique Klein tunneling induced by the anisotropy and tilt of MDF is shown,
which does not depend on the doping levels in \textit{N} and \textit{P} regions of PNJ as the
normal Klein tunneling. To obtain the maximal difference between perfect
transmission direction and the normal direction of PNJ, we analytically
determine the junction direction. In addition, the respective contribution of
anisotropy and tilt underlying the oblique Klein tunneling is also
distinguished, this makes the transmission measurement be useful to reveal the
character of the energy dispersion.

In order to analytically show the oblique Klein tunneling of anisotropic and
tilted MDF, two simplifications for the realistic 8-\textit{Pmmn} borophene
PNJ has been used, i.e., the single Dirac cone and the sharp junction are
considered. The 8-\textit{Pmmn} borophene has two inequivalent Dirac cones
described by low-energy effective Hamiltonian $\hat{H}_{\eta}=\eta\hat{H}_{0}$
with $\eta=\pm$ denoting two cones
\cite{PhysRevB.94.165403,PhysRevB.96.035410,PhysRevB.96.235405}. If neglecting
the intervalley scattering, referring to the detailed presentations for the
$\eta=+$ cone, one can easily derive the results corresponding to $\eta=-$
cone, e.g., the valley-dependent special junction direction $\theta_{m}%
^{\eta}=\eta\theta_{m}$ which may favor the application of 8-\textit{Pmmn}
borophene in valleytronics. The effect of intervalley scattering depends on
the width and direction of PNJ \cite{LogemannPRB2015}. For
the junction width beyond the atomically scale, the intervalley scattering can
be neglected properly due to the large momentum difference between two Dirac
cones \cite{Katsnelson2012}. However, the simplification of sharp junction
implies that the electron wavelength should be larger than junction width,
otherwise a too broad junction will impede the transmission of oblique electron
states across the PNJ \cite{CheianovPRB2006} and is harmful to the application
of PNJ in electron optics \cite{ChenScience2016}. Therefore, our analytical
results are applicable to the low-energy electrons scattered by PNJ with a
rather smooth junction and should be mainly used to understand the novel
physical features accompanying the anisotropic and tilted MDF in contrast to
those of isotropic MDF (e.g., in graphene \cite{Katsnelson2012}). In light of
the experimental advances for confirming different borophene monolayers
\cite{MannixS1513,FengNC2016,PhysRevLett.118.096401}, for fabricating
sharp junctions on the nanoscale \cite{PhysRevB.97.045413}, and for demonstrating
the prominent angular dependence of the transmission probability in planar PNJ
structures \cite{SajjadPRB2012,SutarNL2012,RahmanAPL2015,ChenScience2016}, we
expect the oblique Klein tunneling to be observable in the near future. In order
to compare to future experiments, the further quantitative atomic simulation
is needed by using a proper numerical method \cite{ZhangPRB2017}.

\section*{Acknowledgements}

This work was supported by the National Key R$\&$D Program of China (Grant No.
2017YFA0303400), the NSFC (Grants No. 11504018, No. 11774021, and No. 61504016), the MOST of
China (Grants No. 2014CB848700), and the NSFC program for ``Scientific
Research Center'' (Grant No. U1530401). S.H.Z. is also supported by ``the
Fundamental Research Funds for the Central Universities (ZY1824)" and by ``Chongqing Research Program of Basic Research
and Frontier Technology (cstc2014jcyjA50016)". We
acknowledge the computational support from the Beijing Computational Science
Research Center (CSRC).


\end{document}